\documentclass{PoS}

\title{Gradient flow equation in SQCD}

\ShortTitle{Gradient flow equation in SQCD}

\author{\speaker{Daisuke Kadoh} 
\thanks{This work is supported by JSPS KAKENHI Grant JP19K03853.}\\
Physics Division, National Center for Theoretical Sciences, National Tsing-Hua University, Hsinchu, 30013, Taiwan; 
Research and Educational Center for Natural Sciences, Keio University, Yokohama 223-8521,  Japan\\
        E-mail: \email{kadoh@keio.jp, kado@cts.nthu.edu.tw}}

\author{Naoya Ukita\\
        Center for Computational Sciences, University of Tsukuba, Tsukuba, Ibaraki 305-8577, Japan \\
        E-mail: \email{ukita@ccs.tsukuba.ac.jp} }

\abstract{
We propose a supersymmetric gradient flow in ${\cal N}=1$ SQCD in four dimensions. 
The flow equation is derived in the superfield formalism 
and is also given for component fields of the Wess-Zumino gauge in a gauge covariant manner. 
We find that the flow for the component fields is supersymmetric in a sense that the flow time derivative 
and any supersymmetry transformation commute with each other up to a gauge transformation.
}

\FullConference{37th International Symposium on Lattice Field Theory - Lattice2019\\
		16-22 June 2019\\
		Wuhan, China}

\begin{document}

\section{Introduction}

Gradient flow is widely accepted as a useful tool in various fields of physics. 
In lattice gauge theory, the Yang-Mills flow \cite{Luscher:2010iy,Luscher:2011bx,Luscher:2013cpa} leads 
to many interesting applications 
such as a construction of energy-momentum tensor on a lattice \cite{Suzuki:2013gza, Makino:2014taa}, 
 based on the finiteness of 
flowed field correlators \cite{Luscher:2011bx}. So, if a gradient flow approach is extended to SUSY cases, 
further interesting researches could be created. 

For ${\cal N}=1$ SYM, 
a non-SUSY flow is defined by the YM flow \cite{Luscher:2010iy} and a flow 
for a gaugino \cite{Luscher:2013cpa}, 
which has been studied in \cite{Hieda:2017sqq, Kasai:2018koz, Bergner:2019dim}. 
On the other hand, a SUSY flow is defined by the gradient of the SYM 
action with respect to a vector superfield \cite{Kikuchi:2014rla}. 
The latter flow can be given for the component fields in a gauge covariant 
and supersymmetric manner \cite{Kadoh:2018qwg}, and the finiteness of flowed field correlators 
can be shown for the whole gauge supermultiplet \cite{Kadoh:2018w}. 
\footnote{
See \cite{Kadoh:2019glu} for the SUSY gradient flow in the Wess-Zumino model.}

In this paper, a SUSY gradient flow is derived for ${\cal N}=1$ SQCD in four dimensions.
For the gauge multiplet, a flow equation is defined in the similar manner as that of SYM 
while, for the matter multiplet, a technical modification is needed to define a SUSY flow in the superfield formalism. 
The flow can also be written in the component fields by taking the Wess-Zumino gauge, 
which has the SUSY covariance since the flow time and 
supersymmetry commute with each other up to a gauge transformation.

\section{SQCD in the superfield formalism}

The ${\cal N}=1$ SQCD is a gauge theory that consists of a gauge multiplet $(A_\mu, \lambda,D)$ 
and $N_f $ matter multiplets $(\phi_\pm,q_\pm,F_\pm)_f$ for $f=1,2,\ldots,N_f$. 
Here $A_\mu$ is a gauge field, $\lambda_\alpha, (q_\pm)_\alpha$ are two-component spinors, 
$D,F_\pm$ are real and complex auxiliary fields, respectively.
The gauge group is $SU(N)$ (or $U(N)$) and the generators $T^a$ are hermitian matrices 
normalized as ${\rm tr}(T^a T^b)=\frac{1}{2}\delta_{ab}$. 
Although the gradient flow is a kind of diffusion equation in Euclidean spacetime, 
we derive it in Minkowski spacetime with the superfield formalism.  
The euclidean theory is obtained by the Wick 
rotation $t \rightarrow -it$, $A_0 \rightarrow i A_0$ and replacements of the auxiliary fields $D \rightarrow iD$, $F_\pm \rightarrow i F_\pm$ and
$F^\dag_\pm \rightarrow i F^\dag_\pm$. 
We consider the massless $N_f=1$ case for  simplicity of explanation, and it is straightforward to extend our results to 
a case of multi flavors with a non zero mass term.

The action of ${\cal N}=1$ SQCD in Minkowski space is given by $S_{\rm SQCD}=S_{\rm SYM}+S_{\rm MAT}$: 
\footnote{
We basically follow the notation used in Ref. \cite{Wess:1992cp}. 
The Greek indices $\mu,\nu,\rho,\sigma$ run from $0$ to $3$.
The metric is given by $\eta_{\mu\nu}={\rm diag}\{-1,+1,+1,+1\}$. 
The four component $(\sigma^\mu)_{\alpha\dot\beta}$ and $(\bar\sigma^\mu)^{\dot\alpha\beta}$ 
are defined as $\sigma^\mu=(-i,\sigma^i)$ and $\bar \sigma=(-1,-\sigma^i)$ where
$\sigma^i$ is the standard Pauli matrix. We use $\sigma^{\mu\nu} \equiv \frac{1}{4}(\sigma^{\mu}\bar\sigma^{\nu}-\sigma^{\nu}\bar\sigma^{\mu})$. 
See  \cite{Wess:1992cp} for other details of the notation and useful 
identities for two-component spinors and $\sigma^\mu$ matrix. 
}
\begin{eqnarray}
\label{SYM_action}
&&S_{\rm SYM}=\frac{1}{g^2}\int d^4 x \ 
{\rm tr} \left\{-\frac{1}{2} F_{\mu\nu}^2 
-2 i \bar \lambda \bar\sigma^\mu D_\mu \lambda 
+ D^2 \right\}, \\  
\label{matter_action}
&&S_{\rm MAT}
= \int d^4 x \, \bigg\{
- |D_\mu \phi_+|^2
- |D_\mu \phi_-|^2 + |F_+|^2  + | F_-|^2 \nonumber \\
&&\hspace{2cm} -i \bar q_+\bar\sigma^\mu D_\mu q_+ 
-i q_-\sigma^\mu D_\mu \bar q_- 
+\phi^\dag_+ D\phi_+-\phi_-D\phi^\dag_-  \nonumber \\
&&\hspace{2cm} 
+\sqrt{2} i (
    \phi_+^\dag \lambda q_+ 
+  \phi_- \bar\lambda \bar q_-
 - \bar q_+ \bar\lambda \phi_+ 
-  q_- \lambda \phi^\dag_-
) \bigg\},
\end{eqnarray}
where $F_{\mu\nu} =\partial_{\mu}A_{\nu} -\partial_{\nu}A_{\mu} +i[A_{\mu},A_{\nu}]$ and 
$D_\mu$ is a covariant derivative in the representation of acting fields: 
$D_\mu \varphi =\partial_\mu  \varphi + i [A_\mu,  \varphi]$ 
for $\varphi=\lambda$, $D$, $F_{\rho\sigma}$ and 
$D_\mu X =\partial_\mu  X + i A_\mu X$ for 
$X=\phi_+$, $q_+$, $F_+$, $\phi^\dag_-$, $\bar q_-$, $F^\dag_-$.
The action is invariant under an infinitesimal gauge transformation,
\begin{eqnarray}
\label{normal_gauge_transf}
\delta^g_\omega A_\mu =-D_\mu \omega, \quad
\delta^g_\omega \varphi =i[\omega, \varphi], \quad \delta^g_\omega X = i \omega X, 
\end{eqnarray}
and a SUSY transformation,
\begin{eqnarray}
&&
 \delta_{\xi} A_{\mu} = i \xi \sigma_\mu \bar  \lambda +  i \bar\xi \bar\sigma_\mu \lambda, 
\nonumber \\
&& \delta_{\xi} \lambda =  \sigma^{\mu\nu} \xi F_{\mu\nu} +i \xi D 
\nonumber\\
&& \delta_{\xi} D        =  -\xi \sigma_\mu D_\mu \bar  \lambda +  \bar\xi \bar\sigma_\mu D_\mu \lambda
\nonumber \\
&& \delta_{\xi} \phi_{\pm} =\sqrt{2} \xi q_\pm, 
\label{DF_super}
\\
&& \delta_{\xi} q_{\pm} =\sqrt{2} i \sigma^\mu \bar \xi D_\mu \phi_\pm + \sqrt{2} \xi F_\pm, 
\nonumber \\
&& \delta_{\xi} F_+ = \sqrt{2} i \bar \xi \bar \sigma^\mu D_\mu q_+  +2i \bar \xi \bar \lambda \phi_+,  
\nonumber \\
&& \delta_{\xi} F_- = \sqrt{2} i \bar \xi \bar \sigma^\mu D_\mu q_- -2i \phi_- \bar \xi \bar \lambda,  
\nonumber 
\end{eqnarray}
where $\xi_\alpha$ is a global anti-commuting parameter.

In the superfield formalism,  a superfield $F$ is introduced as a function of 
$z=(x^\mu, \theta_\alpha, \bar\theta_{\dot \alpha})$ which transforms under 
a supersymmetry transformation 
as 
\begin{eqnarray}
\label{super_transf}
\delta F(z)  = (\xi Q +\bar\xi \bar Q) F(z)
\end{eqnarray}
where
 $\theta_\alpha, \bar \theta_{\dot\alpha}$ are two component anti-commuting parameters, and 
$ Q_{\alpha}$ and $\bar Q_{\dot\alpha}$ are differential operators defined by
\begin{eqnarray}
&& 
\label{Q_def}
Q_{\alpha} = 
\frac{\partial}{\partial\theta^{\alpha}} 
-i(\sigma^{\mu})_{\alpha\dot{\alpha}}\bar\theta^{\dot\alpha}\partial_{\mu}, \qquad
  \bar{Q}_{\dot\alpha} = 
-\frac{\partial}{\partial\bar\theta^{\dot\alpha}} 
+i\theta^{\alpha}(\sigma^{\mu})_{\alpha\dot{\alpha}}\partial_{\mu},
\end{eqnarray}
For later use, let us introduce other  differential operators $D_{\alpha}$ and $\bar D_{\dot\alpha}$ as
\begin{eqnarray}
\label{D_def}
D_{\alpha} =
 \frac{\partial}{\partial\theta^{\alpha}}
+i(\sigma^{\mu})_{\alpha\dot{\alpha}}\bar\theta^{\dot\alpha}\partial_{\mu}, \qquad
  \bar{D}_{\dot\alpha} =
 -\frac{\partial}{\partial\bar\theta^{\dot\alpha}}
-i\theta^{\alpha}(\sigma^{\mu})_{\alpha\dot{\alpha}}\partial_{\mu}.
\end{eqnarray}
Note that 
$\{Q_\alpha, \bar Q_{\dot \beta}\}=-\{D_\alpha, 
\bar D_{\dot \beta}\}=2i\sigma^\mu_{\alpha\dot\beta}\partial_\mu$ and 
the other anti-commutation relations are zero.

A superfield $\Lambda$ with a superchiral condition 
$\bar D_{\dot \alpha} \Lambda=0$ is called a chiral superfield and may be expanded as
\begin{eqnarray}
\Lambda(y,\theta) = A(y) + \sqrt{2} \theta   \psi(y)
 + \theta \theta  F(y)  
\end{eqnarray}
with $y^\mu = x^\mu + i \theta \sigma^\mu \bar\theta$. 
Similarly, an anti-chiral superfield $\Lambda^\dag$ is defined by
$D_\alpha \Lambda^\dag=0$.  
A superfield $V$ that satisfies $V=V^\dag$ is called a vector superfield, which is expanded as 
\begin{eqnarray}
&& V(x,\theta,\bar\theta)= \frac{1}{2}{ C (x) } 
+ i\theta  \chi(x) +\frac{i}{2} \theta \theta  
(M(x) +i N(x))
-\frac{1}{2}\theta \sigma^\mu \bar\theta  A_\mu(x) \nonumber \\
&&\qquad 
+ i\theta \theta \bar\theta ( \bar\lambda(x) + 
\frac{i}{2} \bar\sigma^\mu \partial_\mu  \chi(x) )
+\frac{1}{4} \theta \theta \bar\theta\bar\theta 
( D(x)+ \frac{1}{2}  \square C(x)) + h.c. 
\end{eqnarray}
where $C,M,N,A_\mu,D$ are bosonic fields and $\chi,\lambda$ are fermionic fields. 

The SQCD action is then written as
\begin{eqnarray}
\label{superfield_sym_action}
&& S_{SYM} = \frac{1}{2g^2}
\int d^4 x \, {\rm tr} \bigg(\frac{}{}
W^\alpha W_\alpha|_{\theta \theta} + h.c. \bigg), \\
\label{superfield_mat_action}
&& S_{\rm MAT} = 
\int d^4 x \, \bigg\{\frac{}{}
Q_+^\dag e^{2V} Q_+ +  Q_-  e^{-2V} Q_-^\dag \bigg\}
\bigg|_{\theta\theta\bar\theta\bar\theta}, 
\end{eqnarray}
where
\begin{eqnarray}
W_\alpha=-\frac{1}{8} \bar D\bar D e^{-2V} D_\alpha e^{2V}
\end{eqnarray}
with  $V=\sum_{a=1}^{N_c^2-1} V^a T^a$, and $Q_{\pm}$ are chiral superfields,  
\begin{eqnarray}
Q_\pm(y,\theta) = \phi_\pm (y) + \sqrt{2} \theta q_\pm (y)  + \theta\theta F_\pm (y). 
\end{eqnarray}
The superfield action has unwanted $C,\chi,M,N$ fields which are not included in 
eqs.~(\ref{SYM_action}) and (\ref{matter_action}) with an enlarged symmetry. 
Eqs.~(\ref{superfield_sym_action}) and (\ref{superfield_mat_action}) are actually
  invariant under SUSY transformation eq.~(\ref{super_transf})
and an extended gauge transformation generated by a chiral superfield $\Lambda$ as
\begin{eqnarray}
&&e^{2V} \rightarrow  e^{2V^\prime} =e^{2\Lambda^\dag} 
 e^{2V} e^{2\Lambda}, \nonumber \\
 \label{egauge}
&&Q_+  \rightarrow  Q_+^\prime = e^{-2\Lambda} Q_+, \\
&& Q_-  \rightarrow  Q_-^\prime = Q_- e^{2\Lambda}.    \nonumber
\end{eqnarray}
It is possible to remove the extra $C,\chi, M,N$ fields 
by choosing the component fields of $\Lambda$ by hand so that
\begin{equation}
C = \chi = M = N = 0,
\label{WZ_gauge_fixing}
\end{equation}
which is called the  Wess-Zumino gauge. 
With this gauge fixing, Eqs.(\ref{SYM_action}) and (\ref{matter_action}) are reproduced from 
the superfield action eqs.~(\ref{superfield_sym_action}) and (\ref{superfield_mat_action}). 
Then the gauge and SUSY transformations, eqs.~(\ref{normal_gauge_transf}) and (\ref{DF_super}), 
are also reproduced as 
part of enlarged symmetry keeping this gauge.

\section{Derivation of SQCD flow}

In order to define a supersymmetric gradient flow, 
the vector and chiral superfields are transcribed into $t$-dependent superfields where $t$
is a flow time with $V(z,t=0)=V(z)$ and $Q_\pm(z,t)=Q_\pm(z)$ at $t=0$.  
We assume that  all properties of superfields are inherited into the $t$-dependent 
superfields while  $\theta$ and $\bar \theta$ are $t$-independent. 
For instance, supersymmetry transformation is given by  $\delta F(z,t)  = (\xi Q +\bar\xi \bar Q) F(z,t)$ 
where $\xi$ and $\bar\xi$ are $t$-independent and $Q_\alpha, \bar Q_{\dot\alpha}$ defined by 
eq.~(\ref{Q_def}) are kept unchanged.  
In the following, the superfield action  eqs.~(\ref{superfield_sym_action}) and (\ref{superfield_mat_action}) 
are given by replacing $V(z) \rightarrow V(z,t), \ Q_\pm(z) \rightarrow Q_\pm(z,t)$. 

For the gauge multiplet, a supersymmetric gradient flow is defined as
\begin{eqnarray}
\label{V_flow}
\partial_t {V}^a = -\frac{1}{2} g^{ab} \frac{\delta S_{\rm SQCD}}{\delta V^b},  
\end{eqnarray}
where $g_{ab}(V)$ is a metric derived from a norm, 
\begin{eqnarray}
\vert \vert \delta V \vert\vert^2 = \frac{1}{2} \int d^8 z \  {\rm tr} \left(
e^{-2V} \delta e^{2V} e^{-2V} \delta e^{2V} 
\right)(z),
\end{eqnarray}
such that $\vert \vert \delta V \vert\vert^2= \int d^8 z \, g_{ab}(V) \delta V^a \delta V^b$. 
Without the metric, the RHS and the LHS of eq.(\ref{V_flow}) obey different transformation laws
 under the extended gauge transformation which is nonlinear as in the case of general relativity. 
Since this norm is invariant under both of the $t$-independent super and extended gauge transformations, 
the flow equation with the metric eq.~(\ref{V_flow}) is covariant under these two transformations.

Let us consider a supersymmetric flow for the matter multiplets. 
A naive one for $Q_+$ would be given by
$\partial_t Q_+  = \delta S_{\rm SQCD}/ \delta Q_+^\dag$.
This is however not a correct flow equation because
the superchiral condition ($\bar D_{\dot\alpha} Q_+=0$) is not kept 
and the RHS is not proportional to $\square Q_+$ in the free limit as 
\begin{eqnarray}
\label{dSdQ}
\frac{\delta S_{\rm SQCD}}{ \delta Q_+^\dag} =  -\frac{1}{4} DD(e^{2V} Q_+), 
\end{eqnarray}
where the functional derivative is one 
keeping the superchiral condition (See the derivation of field equations in Ref.~\cite{Wess:1992cp}).
If the RHS of eq.~(\ref{dSdQ}) is multiplied by $\bar D^2$, the superchiral condition is kept 
since  $\bar D_{\dot \alpha} \bar D^2=0$, and $\square Q_+$ term appears
as $\bar D^2 D^2=-4 \square$. 
Therefore a SUSY flow for the matter multiplets may be given by 
\begin{eqnarray}
\label{flow_Qp}
&& 
\partial_t Q_+   = -\frac{1}{4} \bar D \bar D 
\left(e^{-2V} \frac{\delta S_{\rm MAT}}{ \delta Q_+^\dag} \right), \\
\label{flow_Qm}
&& \partial_t Q_-   = -\frac{1}{4} \bar D \bar D 
\left( \frac{\delta S_{\rm MAT}}{ \delta Q_-^\dag} e^{2V}\right).
\end{eqnarray}
Note that $S_{\rm MAT}$ is the same as $S_{\rm SQCD}$ under $\delta/\delta Q$ 
since $S_{\rm SYM}$ does not have $Q_\pm$.  
These equations are covariant under $t$-independent 
super and extended gauge transformations. 
We thus find that SQCD flow equations are defined by eqs.~(\ref{V_flow}), (\ref{flow_Qp}), (\ref{flow_Qp}) 
in the superfield formalism.

These flow equations are modified to define SUSY flows for the component fields of the Wess-Zumino gauge.  
The straightforward calculations tell us that 
the Wess-Zumino gauge eq.(\ref{WZ_gauge_fixing})  is not kept
after the time evolution because  
$\partial_t C = -D- (\phi^\dag_+ T^a \phi_+ - \phi_- T^a \phi_-^\dag) T^a \neq 0$ and the RHS of $\partial_t \chi, \partial_tM ,\partial_t N$ are also non zero.
To keep the Wess-Zumino gauge,  
we modify the SUSY flow equations adding an extended gauge transformation as
\begin{eqnarray}
\label{V_flow_mod}
&&\partial_t {V}^a = - \frac{1}{2} g^{ab} \frac{\delta S_{\rm SQCD}}{\delta V^b} \ + \ \delta_\Lambda V^a
 \nonumber \\
\label{flow_Qp_mod}
&& 
\partial_t Q_+  = -\frac{1}{4} \bar D \bar D 
\left(e^{-2V} \frac{\partial S_{SQCD}}{ \partial Q_+^\dag} \right) \ + \ \delta_\Lambda Q_+\\
\label{flow_Qm_mod}
&& \partial_t Q_-  = -\frac{1}{4} \bar D \bar D 
\left( \frac{\partial S_{SQCD}}{ \partial Q_-^\dag} e^{2V}\right) \ + \ \delta_\Lambda Q_-, \nonumber 
\end{eqnarray}
where $\delta_\Lambda$ is an infinitesimal transformation derived from eq.~(\ref{egauge}). 
We take component fields of $\Lambda$ by hand such that $\partial_t C=\partial_t \chi 
=\partial_t M=\partial_t N=0$: 
\begin{eqnarray}
&& A = \frac{D}{2} +\frac{1}{2}(\phi^\dag_+ T^a \phi_+ - \phi_- T^a \phi_-^\dag) T^a, 
\nonumber \\
&& \psi=  -\frac{1}{\sqrt{2}} \sigma^\mu D_\mu \bar\lambda + (\phi^\dag_+ T^a q_+ - q_- T^a \phi_-^\dag) T^a, 
\\
&& F=   (\phi^\dag_+ T^a F_+ - F_- T^a \phi_-^\dag) T^a. \nonumber 
\end{eqnarray}
Using these $A,\psi,F$, the Wess-Zumino gauge fixing is maintained for any nonzero flow time 
as long as
it is set at $t=0$. 

Thus we obtain the flow equations of the components fields: for the gauge multiplet, 
\begin{eqnarray}
&&\partial_t A_\mu = D^\rho F_{\rho \mu} 
-\lambda \sigma_\mu \bar\lambda - \bar\lambda \bar\sigma_\mu \lambda
+ i(\phi_+^\dag T^a D_\mu \phi_+ -D_\mu \phi_-T^a \phi_-^\dag - h.c. ) T^a 
\nonumber \\
&&\hspace{1cm} 
+  (\bar q_+ \bar\sigma^\mu T^a q_+ +q_- \sigma_\mu T^a \bar q_- ) T^a,\\
&&\partial_t \lambda = -\sigma^\mu \bar\sigma^\nu D_\mu D_\nu \lambda
-[\lambda,D] 
-\sqrt{2} \sigma^\mu ( \bar q_+ T^a D_\mu \phi_+  - D_\mu \phi_- T^a \bar q_- ) T^a
\nonumber \\
&&\hspace{1cm} 
-\sqrt{2} i  ( F_+^\dag T^a q_+ -q_- T^a F_-^\dag ) T^a 
- ( \phi_+^\dag \lambda T^a \phi_+ 
+ \phi_- \bar \lambda T^a \phi_-^\dag
+ h.c. ) T^a, \\
&&\partial_t D = D^\mu D_\mu D 
+i ( D_\mu \lambda \sigma^\mu \bar\lambda 
- D_\mu\bar\lambda \bar\sigma^\mu \lambda - h.c.) 
-(\phi_+^\dag DT^a\phi_+ + \phi_- DT^a \phi_-^\dag +h.c.)T^a
\nonumber \\
&&\hspace{1cm}
+ 2 (D^\mu \phi_+^\dag T^a D_\mu \phi_+ -D_\mu \phi_-T^a D_\mu\phi_-^\dag ) T^a 
+ 2\sqrt{2} i (\bar q_+ T^a \bar\lambda \phi_+ + \phi_- \bar\lambda T^a \bar q_- 
-h.c. ) T^a 
\nonumber \\
&&\hspace{1cm} 
+ i (\bar q_+ T^a \bar\sigma^\mu D_\mu q_+
- q_- T^a \sigma^\mu D_\mu \phi_-^\dag - h.c. ) T^a 
-2 ( F_+^\dag T^a F_+ - F_- T^a F_-^\dag ) T^a,
\end{eqnarray}
and, for the matter multiplet, 
\begin{eqnarray}
&&\hspace{-2.8cm}
\partial_t \phi_+ = D^\mu D_\mu \phi_+ +  i\sqrt{2} \lambda q_+ 
- (\phi_+^\dag T^a  \phi_+ - \phi_- T^a \phi_-^\dag) T^a \phi_+, \\
&&\hspace{-2.8cm}
\partial_t q_+ = -\sigma^\mu \bar\sigma^\nu D_\mu D_\nu q_+ 
+i \sqrt{2} \lambda F_+ -\sqrt{2} \sigma^\mu \bar\lambda D_\mu \phi_+ -Dq_+
\nonumber
\\
&&\hspace{-1.5cm} 
-( \phi^\dag_+ T^a \phi_+  - \phi_- T^a \phi^\dag_- ) T^a q_+
-2 ( \phi_+^\dag T^a q_+ - q_- T^a \phi_-^\dag ) T^a \phi_+, \\
&&\hspace{-2.8cm}
\partial_t F_+ = D^\mu D_\mu F_+ - 2 D F_+ 
+\sqrt{2} (D_\mu\bar\lambda \bar\sigma^\mu q_+
- \bar\lambda \bar\sigma^\mu D_\mu q_+) -2 \bar\lambda \bar\lambda \phi_+
\nonumber
\\
&&\hspace{-1.5cm} 
-( \phi^\dag_+ T^a \phi_+  - \phi_- T^a \phi^\dag_- ) T^a F_+
+2 ( \phi_+^\dag T^a q_+ - q_- T^a \phi_-^\dag ) T^a q_+
\nonumber
\\
&&\hspace{-1.5cm} 
-2 ( \phi_+^\dag T^a F_+ - F_- T^a \phi_-^\dag ) T^a \phi_+,
\end{eqnarray}
where similar equations are given for $\phi_-,q_-,F_-$. 
Note that we are now working on the Minkowski space. After the Wick rotation, 
the euclidean flow equations, which are kinds of diffusion equations, are obtained for all fields.

These equations are gauge covariant under the $t$-independent gauge transformations. 
The supersymmetry transformations for the  flowed fields are defined by
 eq.~(\ref{DF_super})
replacing all of the fields with the corresponding $t$-dependent fields. 
Then one can show that 
\begin{eqnarray}
[\partial_t,\delta_\xi]=\delta^g_{\omega},   \quad \omega \equiv - i D_\mu (\xi \sigma_\mu \bar  \lambda +  \bar\xi \bar\sigma_\mu \lambda),  
\label{consistency_relation}
\end{eqnarray}
where $\delta^g_\omega$ is an infinitesimal gauge transformations with a gauge parameter 
$\omega$, which is given by eq.~(\ref{normal_gauge_transf}).
The obtained flows are supersymmetric since the RHS of eq.~(\ref{consistency_relation}) 
vanishes for gauge invariant quantities.

These flow equations can be simplified without breaking the supersymmetry covariance. 
The gradient of $S_{\rm SYM}$ is also employed to define a flow for the gauge multiplet in eq.~(\ref{V_flow}) 
instead of $S_{\rm SQCD}$. 
Although the flows for the matter multiplet  are given by  the same expressions as eqs.~(\ref{flow_Qp}) and (\ref{flow_Qm}), this change affects the whole of flow equations 
for the component fields because $\Lambda$ changes. 
A straightforward calculation yields 
\begin{eqnarray}
&&\partial_t A_\mu = D^\rho F_{\rho \mu} 
-\lambda \sigma_\mu \bar\lambda - \bar\lambda \bar\sigma_\mu \lambda, \\
&&\partial_t \lambda = -\sigma^\mu \bar\sigma^\nu D_\mu D_\nu \lambda
-[\lambda,D],\\
&&\partial_t D = D^\mu D_\mu D 
+i ( D_\mu \lambda \sigma^\mu \bar\lambda 
- D_\mu\bar\lambda \bar\sigma^\mu \lambda - h.c.),
\end{eqnarray}
and 
\begin{eqnarray}
&&\partial_t \phi_+ = D^\mu D_\mu \phi_+ +  i\sqrt{2} \lambda q_+, \\
&&\partial_t q_+ = -\sigma^\mu \bar\sigma^\nu D_\mu D_\nu q_+ 
+i \sqrt{2} \lambda F_+ -\sqrt{2} \sigma^\mu \bar\lambda D_\mu \phi_+ -Dq_+, \\
&&\partial_t F_+ = D^\mu D_\mu F_+ - 2 D F_+ 
+\sqrt{2} (D_\mu\bar\lambda \bar\sigma^\mu q_+
- \bar\lambda \bar\sigma^\mu D_\mu q_+) -2 \bar\lambda \bar\lambda \phi_+. 
\end{eqnarray}
where similar equations are given for $\phi_-,q_-,F_-$.
These simplified flows also satisfy eq.~(\ref{consistency_relation}).

\section{Summary and future outlook}

We have derived a supersymmetric flow equation in four-dimensional ${\cal N}=1$ SQCD 
using the superfield formalism. The flows can also be written 
in the component fields of the Wess-Zumino gauge in a gauge covariant manner.
We have directly confirmed that the obtained flow is supersymmetric 
in a sense that the flow time derivative and supersymmetry commute 
with each other up to a gauge transformation at any non-zero flow time.

As a next step of our research, we are investigating the finiteness of flowed field correlators 
at all order of the perturbation theory.  Then we aim to construct the Ferrara--Zumino multiplet 
on the lattice using a small flow time expansion of flowed field correlators,  after which we will 
use our method in actual numerical simulations to study Seiberg duality in ${\cal N}=1$ SQCD. 

SUSY flows of ${\cal N}=2$ and ${\cal N}=4$ theories would be derived using similar techniques 
given in this paper. If they are given in terms of ${\cal N}=1$ superfield formalism, 
eq.~(\ref{consistency_relation}) would not be satisfied for the whole of extended supersymmetry. 
In the case, it is not obvious how to hold the finiteness of flowed field correlators. 
Therefore further studies are needed to understand SUSY flows 
in extended SUSY theories.


\end{document}